\documentclass[12pt]{article}
\usepackage{epsfig}

\def\beq{\begin{equation}}
\def\eeq#1{\label{#1}\end{equation}}
\def\eeqn{\end{equation}}

\def\beqa{\begin{eqnarray}}
\def\eeqa#1{\label{#1}\end{eqnarray}}
\def\eeqan{\end{eqnarray}}

\let\bar=\overbar

\def\Dslash{\not{\hbox{\kern-4pt $D$}}}
\def\dslash{\not{\hbox{\kern-2pt $\del$}}}

\def\msb{{\bar{\ssstyle M \kern -1pt S}}}

\def\Title#1{\begin{center} {\Large {\bf #1} } \end{center}}

\begin{document}

\Title{Jet physics at HERA, Tevatron and LHC}

\bigskip\bigskip


\begin{raggedright}  

{\it Christophe Royon\index{Royon, C.}\\
IRFU-SPP, 
CEA Saclay,
F91 191 Gif-sur-Yvette, France}
\bigskip\bigskip
\end{raggedright}

\section{Introduction}
\begin{figure}[htb]
\begin{center}
\epsfig{file=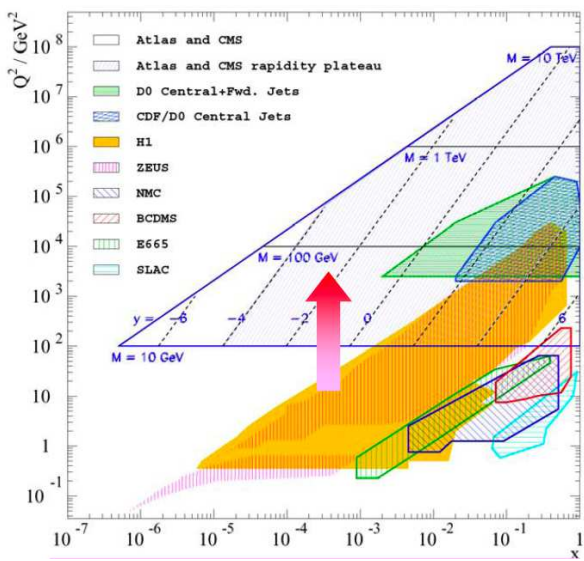,height=2.9in}
\caption{Kinematical domain reached by the experiments at HERA, Tevatron and LHC.}
\label{kin}
\end{center}
\end{figure}

In this short report, we discuss the Jet Physics results and perspectives
at HERA, Tevatron and LHC. The different accelerators are complementary 
as shown in Fig.~\ref{kin}, where the kinematical plane in $(x,Q^2)$ is
displayed ($x$ and $Q^2$ are respectively the proton momentum fraction
carried by the interacting parton and the transferred energy squared carried
by the virtual photon). HERA allows to reach very low values of
$x$ at low $Q^2$ ($x\sim 10^{-6}$), whereas the Tevatron (and the LHC)
very high values of $Q^2$ at high $x$ ($Q^2\sim 3~ 10^5$, $10^8$ GeV$^2$
at the Tevatron and the LHC respectively). In the following, we will benefit
from the differences between the accelerators to assess the proton structure 
in a wide kinematical domain.

We will start this report by describing the constraints
on the proton structure (quark and gluon densities) using inclusive jets at HERA
and the Tevatron. The study of the mutijet cross sections will be discussed in
the second part of the report since it is a fundamental topic for the LHC and
the searches for new particles in the jet channels. Another background related
to SUSY and Higgs boson searches is the $W+b$ jet and $Z+b$ jet events and we
will give the most recent results from the Tevatron. We will finish the report
by describing the low $x$ dynamics which can be probed in forward jets at HERA
and Mueller-Navelet jets at the Tevatron/LHC in particular.

\section{Inclusive jets at HERA and the Tevatron}

\subsection{High $Q^2$ jet measurements at HERA}
\begin{figure}
\begin{center}
\epsfig{file=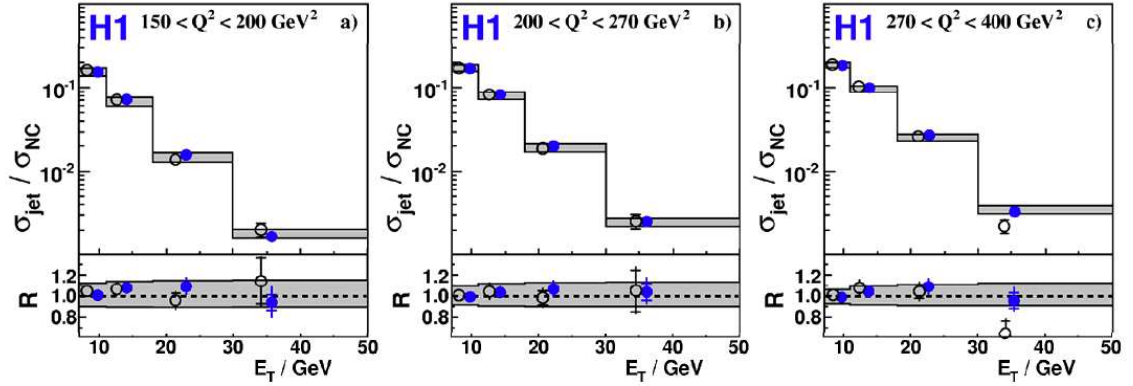,height=2.3in}
\caption{Ratios of the jet production to the neutral current cross sections as a
function of jet $E_T$ in three different $Q^2$ regions.}
\label{highq2}
\end{center}
\end{figure}

\begin{figure}
\begin{center}
\epsfig{file=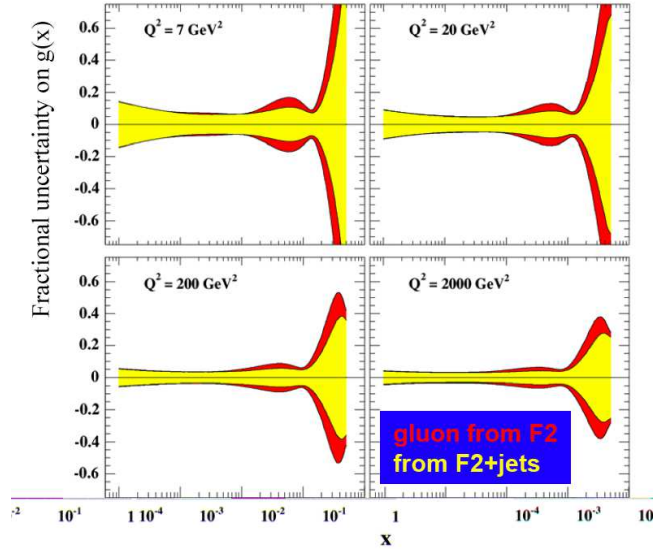,height=2.9in}
\caption{Fractional uncertainty on gluon density in the proton in four different $Q^2$ bins determined
using the proton structure function $F_2$ data measured at HERA (in red) and 
the jet cross sections in addition (in yellow).}
\label{gluon}
\end{center}
\end{figure}

In addition to the measurement of the proton structure function $F_2$ which
allows to access directly the structure of the proton in terms of quarks and
gluons, it is possible to probe the gluon density at high $x$ using jet
measurements at HERA. The H1 and ZEUS collaborations at HERA measured the ratios of the
jet and neutral current cross sections~\cite{heraratios} to remove many systematic uncertainties
as shown in Fig.~\ref{highq2}.
The jet cross section measurement allows to perform a direct test of the
next-to-leading order (NLO) QCD evolution, and allows to constrain the parton
distribution functions (PDF) and the values of $\alpha_S$. The effect of
including or not the jet cross sections in addition to the proton
structure function measurements to constrain further the parton
density at high $x$ in the proton is shown in Fig.~\ref{gluon}. The
uncertainties on the gluon density at high $x$ are still very large (typically
larger than 20\% for $x>0.3$ at high $Q^2 \sim 2000$ GeV$^2$, increasing to
100\% at low $Q^2$), and we will study if the Tevatron (and then the LHC) can
reduce this uncertainty further.

The H1 and ZEUS collaborations also measured the charged current jet production
cross section for jet transverse energies above 100 GeV. A good agreement is
found with NLO calculations but in addition to PDF uncertainties, there is a
large theoretical uncertainty at high $x$ which shows the need for NNLO
calculations~\cite{charged}.

\subsection{Inclusive jet cross section measurements at the Tevatron}

The inclusive jet cross section measurements at the Tevatron rely on the
determination of the jet
energy calibration, which leads to the largest systematic uncertainties.
Jet measurements are corrected either to particle level or to parton
level, depending on the measurements and the collaboration. Jet measurements are
performed using either a cone or the $k_T$ algorithm.
The jet energy scale
is determined mainly using $\gamma+$jet events. In the D0 collaboration, the
corrected jet energy is obtained using the following method
\begin{eqnarray}
E_{jet}^{corr} = \frac{E_{jet}^{uncorr} - Off}{Show \times Resp}
\end{eqnarray}
where $E_{jet}^{corr}$ and $E_{jet}^{uncorr}$ are the corrected and uncorrected
jet energies respectively. The offset corrections ($Off$) are related to uranium noise
and pile-up and are determined using zero-bias data. The showering corrections
($Show$) take into account the energy emitted outside the jet cone because of
the detector and dead material and, of course, not the
physics showering outside the jet cone which corresponds to QCD radiation
outside the cone. The jet response ($Resp$) is the largest correction, and can
be subdivided in few corrections. The first step is to equalize the calorimeter
response as a function of rapidity, and the jet response is then measured for
the central part of the calorimeter only using the $p_T$ balance in $\gamma+$jet
events. Some additional small corrections related to the method biases are
introduced. One important additional correction deals with the difference in
response between quark and gluon jets. The difference was studied both in data
and in Monte Carlo (using for instance the $\gamma+$jet and the dijet samples
which are respectively quark and gluon dominated) and leads to a difference of
4 to 6\% as a function of jet $p_T$, which is not negligible if one wants a
precision on jet energy scale of the order of 1\%. This has an important
consequence. The jet energy scale is not universal but sample dependent. QCD
jets (gluon dominated) will have a different correction with respect to the $t
\bar{t}$ events for instance which are quark dominated. The CDF collaboration
follows a method which is more Monte Carlo oriented using beam tests and single
pion response to tune their Monte Carlo. At the LHC, it will be possible to use
$Z+$jets which do not suffer from the ambiguity of photon identification in the
detector.

The uncertainties reached by the D0 collaboration concerning the determination
of jet energy scale are of the order of 1.2\% for jet $p_T$ between 70-400 GeV
and in a wide range of rapidity around zero (the uncertainty is of the order of 2\% for a
rapidity of 2.5). This allows to make a very precise measurement of the jet
inclusive cross section as a function of their transverse momentum.

\begin{figure}[htb]
\begin{center}
\epsfig{file=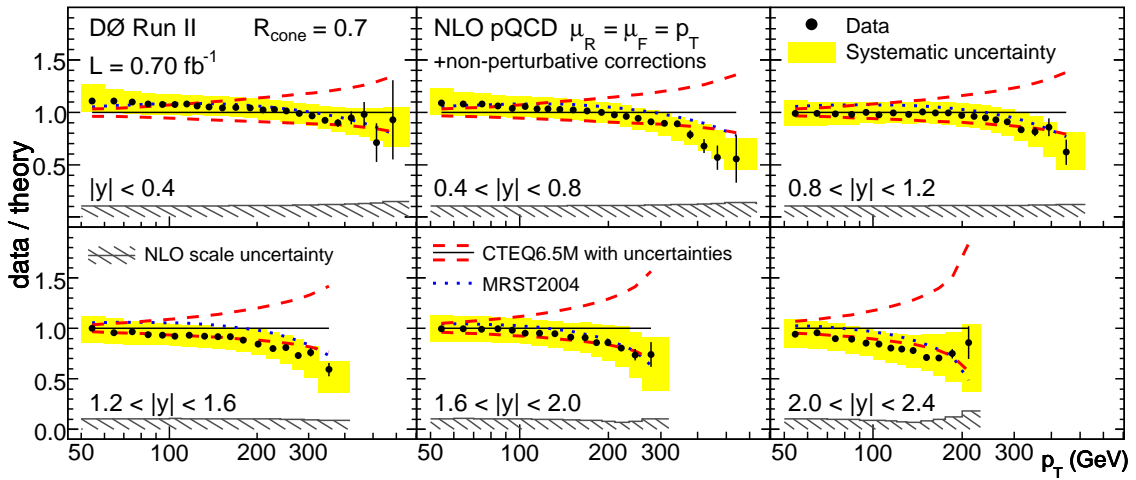,height=2.5in}
\caption{Data over theory for the inclusive $p_T$ cross section measurement for
the D0 collaboration using the 0.7 jet cone. Data are compared to NLO QCD calculations using the
CTEQ6.5M parametrisation.}
\label{inclusive1}
\end{center}
\end{figure}

\begin{figure}[htb]
\begin{center}
\epsfig{file=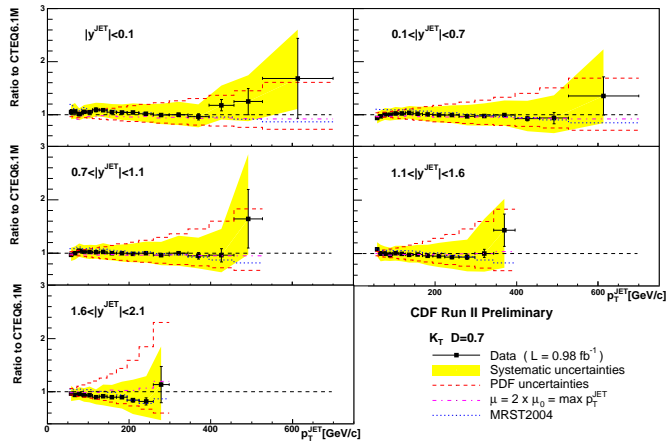,height=2.8in}
\caption{Data over theory for the inclusive $p_T$ cross section measurement for
the CDF collaboration using the $k_T$ algorithm. Data are compared to NLO QCD calculations using the
CTEQ6.1 parametrisation.}
\label{inclusive2}
\end{center}
\end{figure}

The measurement of the inclusive jet cross section~\cite{inclusive} was performed by the D0 and
CDF collaborations at the Tevatron using a jet cone algorithm with a cone size
of 0.7 (D0 and CDF) and the $k_T$ algorithm (CDF). Data are corrected to hadron
level (D0) or parton level (CDF). The motivation of this measurement is double:
it is sensitive to beyond standard model effects such as quark substructure and
to PDFs, especially the gluon density at high $x$. Historically, the excess
observed by the CDF collaboration in 1995 concerning the inclusive jet $p_T$
spectrum compared to the parametrisations was suspected to be a signal of quark
substructure but it was found that increasing the gluon density at high $x$
could accomodate these data. This raises the question of PDFs versus beyond
standard model effects, and the interpretation of data in general.
Data are compared with NLO QCD calculations
using either CTEQ6.5M~\cite{cteq} for D0 or CTEQ6.1 for CDF (the uncertainties of the
CTEQ6.5M parametrisation are two times smaller). A good agreement is found over
six orders of magnitude. The ratio data over theory for the D0 and CDF
measurements are given in Figs.~\ref{inclusive1} and \ref{inclusive2}. A good
agreement is found between NLO QCD and the D0 or CDF measurements with a
tendency of the CTEQ parametrisation to be slightly lower than the data at high
jet $p_T$. The MRST2004~\cite{cteq} parametrisation follows the shape of the measurements.
Given the precision obtained on jet energy scale, the uncertainties obtained 
by the D0 collaboration are lower than the PDF ones and will allow to constrain
further the PDFs (the uncertainties of the CDF collaboration are about two times
larger). The D0 collaboration took also special care of the uncertainty
correlation studies, by giving the effects of the 24 sources of systematics in
data.
 
In addition, the CDF collaboration measured the dijet mass cross
section~\cite{dijetmass} above
180 GeV, and up to 1.2 TeV. No excess was found with respect to NLO QCD
calculations and this measurement allows to exclude excited quarks below 870
GeV, $Z'$ (resp. $W'$) below 740 (resp. 840) GeV~\footnote{Stronger limits 
on $W'$ and $Z'$ mass limits come from lepton based searches}, 
and technirho below 1.1 TeV.

The question rises if PDFs can be further constrained at the LHC using inclusive
measurements. The PDF uncertainties are typically of the order of 15\% for a jet
$p_T$ of 1 TeV, and 25\% of 2 TeV for $1<|\eta_{jet}|<2$ (without taking into
account the new Tevatron measurements which we just discussed). A typical
uncertainty of 5\% (resp. 1\%) on jet energy scale leads to a systematic
uncertainty on 30 to 50\% (resp. 6 to 10\%) on the jet cross section. A precise
determination of the jet energy scale at the LHC will thus be needed to get
competitive measurements at the LHC.

\subsection{How do PDF uncertainties affect LHC potential?}
Another question to be raised is to know whether the uncertainty on PDFs (and
also of higher order effects) can affect the LHC discovery potentual. As an
example, let us consider the Higgs boson production. The cross sections are
known precisely both for background and signal (typically the uncertainties on
$\sigma(gg
\rightarrow H)$ and on $\sigma(qq \rightarrow H qq)$ cross sections due to 
PDFs are 
respectively less than 5 and 15\% over the full Higgs boson
mass range). However, there are additional uncertainties related to higher order
effects. For example, for Higgs production for a Higgs mass of 120 GeV, NNLO
effects are of the order of 9\% (for $Z$ production, it is of the order of 4\%).
Both sets of uncertainties have to be taken into account in the predictions.

On the other hand, the LHC potential can be affected if the background is poorly
known. PDF uncertainties can thus have an impact on searches (extra dimensions,
single top, SUSY...). As an example, we can quote the search for $qqqq$ contact
interactions for a given compactification scale which can appear as an excess in
the dijet mass spectrum. For a compactification scale of 2 TeV, and 2 extra
dimensions, the effect of contact interactions is found to be of the same order
as the present PDF uncertainties.

\section{Multijet cross section measurements at the Tevatron and at HERA}

The measurement of multijet cross sections at the Tevatron and at HERA (and
later on at the LHC) is fundamental to constrain the PDFs and to tune the Monte
Carlo, since it is a direct background entering in many searches for Higgs
bosons or new particles at the LHC. We can quote for instance the search for
Higgs bosons in association with $t \bar{t}$, the measurement of the $t \bar{t}$
production cross section, the search for $R$-parity violated SUSY (which can
lead up to 8-10 jets per event...).

\subsection{Measurement of $\Delta \Phi$ between jets in D0}
The advantage of the measurement of the difference in azimuthal angle 
between two leading jets in an inclusive QCD sample as was performed in D0 is
that 
there is no need of precise knowledge of jet
energy scale (the measurement is dominated by the knowledge of jet angles). 
The $\Delta \Phi$ spectrum was measured in four
different regions in maximum jet transverse momentum, and a good agreement was
found with NLO calculations except at very high $\Delta \Phi$ where soft
radiation is missing~\cite{deltaphi}. PYTHIA~\cite{pythia} shows a disagreement at small
$\Delta \Phi$, showing a lack of initial state gluon radiation, while
HERWIG~\cite{herwig} shows a good agreement with data.

\subsection{Measurement of multijet and $\gamma +$jet cross sections}
The H1 and ZEUS collaborations measured the 2 and 3 jet production cross section
relatively to the neutral current one to reduce systematics. A good agreement is
found with NLO calculations~\cite{multijethera}.

The D0 collaboration measured the inclusive production of isolated $\gamma +$
jets in different detector regions requiring a central photon and a central or a
forward jet. It distinguished the cases when the photon and
the jet are on the same or opposite side. The cross section has been found in
disagreement with NLO QCD expectations both in shape and normalisation and the
reason is unclear~\cite{gammajet}.

\subsection{Jet shape measurements in CDF}
The jet shape is dictated by multi-gluon emission from primary partons, and is
sensitive to quark/gluon contents, PDFs and running $\alpha_S$, as well as
underlying events. We define $\Psi$ which is sensitive to the way the energy is
spread around the jet center
\begin{eqnarray}
\Psi (r) = \frac{1}{N_{jets}} \Sigma_{jets} \frac{P_T(0,r)}{P_T^{jet}(0,R)}
\end{eqnarray}
where $R$ is the jet size.
The energy is more concentrated towards the jet center for quark than for gluon
jets since there is more QCD radiation for gluon jets (which
means that $\Psi$ is closer to one for quark jets when $r \sim 0.3 R$ for
instance. The CDF collaboration measured $\Psi(0.3/R)$ for jets with $0.1 < |y|
< 0.7$ as a function of jet $p_T$ and found higher values of $\Psi$ at high
$p_T$ as expected since jets are more quark like~\cite{shape}. This measurement also helps
tuning the PYTHIA and HERWIG generators since it is sensitive to underlying
events in particular.

\begin{figure}[htb]
\begin{center}
\epsfig{file=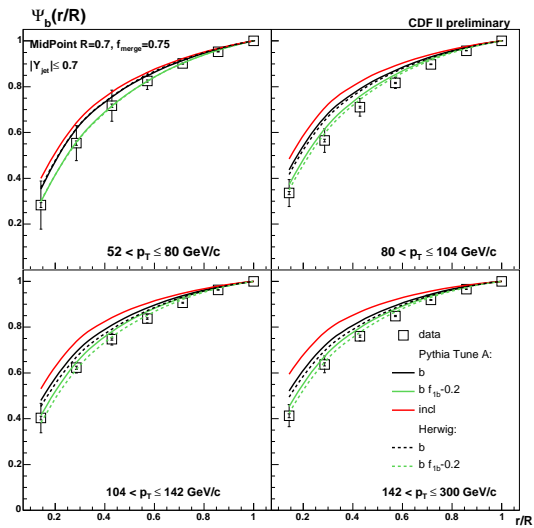,height=2.8in}
\caption{Measurement of the $b$-jet shapes and comparison with the predictions
of the PYTHIA and HERWIG Monte Carlo (see text).}
\label{bjet}
\end{center}
\end{figure}

The CDF collaboration also studied the jet shapes for $b$-jets in four different
$p_T$ bins~\cite{shapeb}, and the result
is given in Fig.~\ref{bjet}. The default PYTHIA and HERWIG Monte Carlo in black
full and dashed lines respectively are unable to describe the measurement.
Compared to the inclusive jet shape depicted in Fig.~\ref{bjet} in full red line
for PYTHIA, the tendency of the $b$-jet shape is definitely the right one,
leading to smaller values of $\Psi$ as expected, but the measurement leads to a
larger difference. The effect of reducing the single $b$-quark fraction by 20\%
leads to a better description of data as it shown in green in Fig.~\ref{bjet}.
The fraction of $b$-jets that originate from flavour creation (where a single
$b$-quark is expected in the same jet cone) over those that originate from
gluon splitting (where two $b$-quarks are expected in the same jet cone)  is
different in Monte Carlo and data.

The CDF collaboration also measured the $b \bar{b}$ dijet cross section as a
function of the leading jet $p_T$ and the difference in azimuthal angle between
the two jets and it leads to the same conclusion, namely that PYTHIA and HERWIG
underestimates the gluon splitting mechanism~\cite{dijetmass}.

\section{Underlying events at Tevatron and LHC}
The CDF collaboration measured underlying events at the Tevatron and used these
measurements to tune in particular the PYTHIA generator. $pp$ or $p \bar{p}$
interactions are namely not as simple as interactions in $ep$ colliders. In
addition to the hard scattering producing dijets, high $p_T$ leptons...,
spectator partons produce additional soft interactions called underlying events.
The main consequence is that it introduces additional energy in the detector not
related to the main interaction which need to be corrected. 

To study this kind of events, the idea is quite simple. It is for instance
possible to use dijet events and we can distinguish in azimuthal angle three
different regions: the ``toward" region around the leading jet direction defined
by a cone of 60 degrees around the jet axis, the ``away" region in the opposite
direction to the jet, and the ``transverse" region the remaining regions far
away from the jet and the ``away" region. In dijet events, the ``transverse"
region will be dominated by underlying events. The CDF collaboration measured
the charged multiplicity and the charged transverse evergy as a function of jet
transverse energy and used these quantities to tune the PYTHIA Monte Carlo
leading to the so called Tune A and Tune AW~\cite{dijetmass}. 

\begin{figure}[htb]
\begin{center}
\epsfig{file=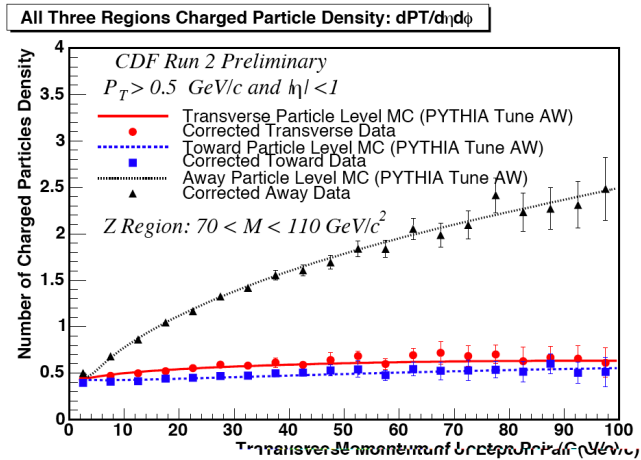,height=2.5in}
\caption{Measurement of the charged particle density for Drell Yan events in the
``toward", ``away" and ``transverse" regions compared to PYTHIA Tune AW.}
\label{underlying}
\end{center}
\end{figure}

Clean Drell Yan events can also be used to tune underlying
events~\cite{dijetmass}. The lepton
pair defines the ``toward" region while the ``away" and ``transverse" regions
are defined in the same way as for dijets. As an example, we give in
Fig.~\ref{underlying} the charged particle density as a function of the
transverse momentum of the lepton pair in the three regions compared with the
Tune AW of PYTHIA.

At the LHC, one of the first measurements to be performed will be related to the
tuning of underlying events in the generators. Present tunings between the 
different Monte Carlo (PYTHIA, PHOJET, HERWIG) show differences
up to a factor six concerning the average multiplicity of charged particles as a
function of the $p_T$ of the leading jet as an example, and it is crucial to
tune the Monte Carlo to accomplish fully the LHC program.

\section{Measurements of the $W+$jet and $Z+$jet cross sections at the Tevatron}

The measurements of the $W+$jet and $Z+$jet cross sections are specially
important since they are a background for many searches and especially the
search for the Higgs boson.

\subsection{Measurements of the $W+X$ cross sections}
The D0 collaboration measured the ratio of the $W+c$ to the inclusive cross
section 0.074 $\pm$ 0.019 (stat.) $\pm^{0.012}_{0.014}$ (syst.) in agreement
with NLO calculation~\cite{wcharm}. It will be important to redo this measurement with higher
statistics since it is directly sensitive to the $s$-quark PDF.

The $W+X$ cross section measurement at the LHC is considered to be one of the
``standard" candles with small theoretical uncertainties (the NNLO scale
dependence is less than 1\%) and could be used even for luminosity measurements.
Unfortunately, the PDFs are not so well known in the kinematical region where
the $W+X$ cross section is measured. The average value of $x$ ($<x> \sim
7.10^{-3}$ with $5.10^{-4}<x<5.10^{-2}$) is not in the valence region and thus
not in the region where quarks are best known. The differences between PDFs lead
to an uncertainty on the $W+X$ cross section of the order of 8\% which is not
precise enough to be used as a luminosity monitor. An independant better
determination of the PDFs would change the conclusions.

\subsection{Measurement of the $Z+b$ and $W+b$ cross sections}
The motivation to measure the $Z+b$-jet cross section is quite clear: this is a
direct background for Higgs boson searches and it is also sensitive to the $b$
quark content of the proton. The measurements of the $Z+b$-jet 
and $W+b$-jet cross sections were
performed by the CDF collaboration at the Tevatron $\sigma(Z+b~jets)=$0.86 $\pm$ 
0.14 $\pm$ 0.12 pb and $\sigma(W+b-jets) \times BR(W \rightarrow l \nu) =
2.74 \pm 0.27 (stat.) \pm 0.42 (sys.)$ pb in agreement with NLO calculations and
PYTHIA predictions~\cite{zbwb}. The CDF collaboration also compared the differential
distributions in jet $p_T$ and rapidity as an example and the distributions are
found in good agreeement with PYTHIA.

\section{Forward jets and Mueller Navelet jets}

\subsection{Low $Q^2$ jets at HERA}
We discussed so far only high $E_T$ jets at high $Q^2$ and the question raises
about what happens at low $Q^2$ and how low in $Q^2$ and jet $p_T$ is
perturbative QCD at NLO reliable. In other words, BFKL~\cite{bfkl} effects are supposed to
appear at very low $Q^2$. The H1 collaboration measured the inclusive jet cross
section differentially in $Q^2$ ($d\sigma / dQ^2$) for jet $p_T$ greater than 5
GeV and a discrepancy of about a factor 2 between NLO calculations and the 
measurement is found for $Q^2 \sim 6$ GeV$^2$. The reason can be due to missing
higher order effects (NNLO) or missing low $x$ resummation terms present in the
BFKL equation~\cite{lowq2h1}.

To test further the low $x$ dynamics, the H1 and ZEUS collaborations measured
forward jet production cross sections. The idea is simple: we ask jets to be
emitted in the ``forward" region, as far as possible in rapidity from the
scattered electron. When the jet $p_T^2$ and the virtual photon $Q^2$ are close,
the DGLAP NLO cross section~\cite{dglap} is expected to be small because of the $k_T$
ordering of the partons in the ladder in the DGLAP evolution. The BFKL cross
section is expected to be much higher since there is no $k_T$ ordering of the
emitted gluons. The kinematical region probed by the H1 collaboration is 
$10^{-4}<x<4.10^{-3}$, $p_{T}(jet)>3,5$ GeV,
$7<\theta_{jet}<20$ degrees, $0.5< p_T^2/Q^2<5$ to enhance the BFKL resummation
effects~\cite{fwdjet}. A discrepancy between NLO QCD prediction and the measurement is found
on the differential forward jet $d \sigma/dx$ cross section at low $x$ (the
discrepancy is about a factor 3 for $x \sim 0.0005$. The H1 collaboration also
looked at the production cross section of two forward jets and one central jet
and some discrepancy is found again at low $x$. 

\begin{figure}[htb]
\begin{center}
\epsfig{file=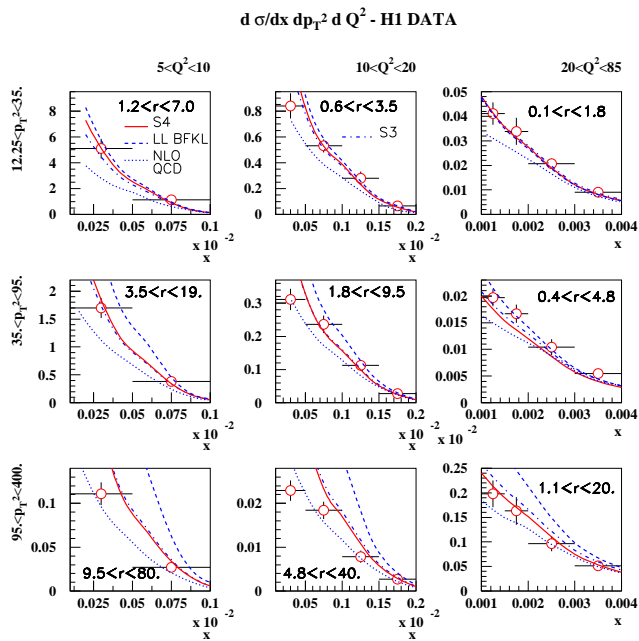,height=3.5in}
\caption{Triple differential cross section measured by the H1 collaboration.}
\label{forward}
\end{center}
\end{figure}

To study further how one moves
from the BFKL dynamics to the DGLAP one, the H1 collaboration measured the
triple differential jet cross section $d \sigma/dx dp_T^2 dQ^2$~\cite{fwdjet} as a function of $x$
for different regions in $Q^2$ and $p_T^2$. The measurement is shown in
Fig.~\ref{forward}~\cite{ourfwd}. The NLO QCD prediction is displayed in dotted line and
describes the cross section at high $p_T$ but not at low $p_T$ where it
undershoots the data. The LL BFKL prediction leads to a good description at low
$p_T$ (or in the case when $r=p_T^2/Q^2$ is close to 1 as expected since BFKL
effects are dominant in this kinematical region, and overshoots the data at high
$p_T$. BFKL NLL leads to a good description of data over the full range. In
Fig.~\ref{forward}, we display two different resummation schemes for BFKL NLL
called S3 and S4 which both lead to a good description~\cite{ourfwd}. It is worth noticing
that implementing the higher-order corrections in the impact factor due to exact
gluon kinematics in the $\gamma^* \rightarrow q \bar{q}$ transition improves
further the description of data~\cite{ourfwd}. This measurement shows a clear discrepancy with
DGLAP NLO calculation and is well described by the NLL BFKL formalism, and it
would be nice to know the effects of higher orders corrections of the DGLAP
prediction.

The ZEUS collaboration also studied the forward jet cross section. They measure
the 3 jet cross section and they see a disagreement with NLO QCD when the jets
are in the forward region~\cite{fwdjet}.

\subsection{Mueller Navelet jets at the Tevatron and the LHC}
The same idea as the forward jets at HERA can be used at the Tevatron and the
LHC. Mueller Navelet jets are jets produced in $pp$ and $p \bar{p}$ collisions,
requiring these two jets to be as far away as possible in rapidity, and to have
about the same transverse momentum. For the same reason as for forward jets, the
$k_T$ ordering of the gluons of the ladder ensures that the DGLAP cross section
is low whereas the BFKL one is expected to be higher. Another easier observable
is the measurement of the difference in azimuthal angle between the two forward
jets. Since there are few gluons emitted for the DGLAP evolution, the $\Delta
\Phi$ value is peaked towards $\pi$ whereas the BFKL expectation will be a
flatter distribution in $\Delta \Phi$ because of the emitted gluons. This
measurement can be performed at the Tevatron and the LHC and can be a test of
BFKL resummation effects~\cite{mnjets}.

\section{Conclusion}
In this short report, we presented many new results from HERA and the Tevatron
concerning jet physics and also some expectations for the LHC. In particular,
the new measurement of the inclusive jet cross section at the Tevatron is
complementary to the HERA jet cross section measurements and is fundamental to
constrain further the gluon density at high $x$, which is useful for searches at
the LHC in the jet channel, especially for a better knowledge of background. The
multijet cross section measurements is also in agreement with NLO QCD
calculations and is also fundamental for the LHC. The $\gamma +$jet cross
sections is in discrepancy with NLO calculation and the reason is unclear. The
$W+$jet and $Z+$jet cross sections are in general in agreement with NLO
calculations but the uncertainties are still large and will benefit from higher
statistics. We finished the report by describing the forward jet and Mueller
Navelet jet measurements which are senstive to low $x$ resummation effects given
by the BFKL equation. Many other topics such as diffraction and the search for
diffractive exclusive events in the jet channel by the CDF collaboration, and
the implications for the LHC diffractive program were not described because of
lack of time~\cite{final}

\def\Discussion{
\setlength{\parskip}{0.3cm}\setlength{\parindent}{0.0cm}
     \bigskip\bigskip      {\Large {\bf Discussion}} \bigskip}
\def\speaker#1{{\bf #1:}\ }
\def\endDiscussion{}

\end{document}